# Incompressible Canonical Quantization

C P Viazminsky
Department of Physics
University of Aleppo
Aleppo, Syria
Kayssarv@hotmail.com

## Abstract

The notion of incompressible momentum observables is introduced. It is shown that when the metric in a manifold has a certain form, a set of canonically conjugate variables $X_k$ and $\hat{P}_k$, in which $\hat{P}_k$ are incompressible, can be constructed. Based on this set of variables, the quantum mechanical description of the motion of a particle in a manifold, is identical to that associated with the familiar canonically conjugate variables $x_k, \hat{p}_k$ in an Euclidean space with Cartesian coordinates. The controversy related to non-uniqueness of momentum operators when the range of a coordinate is a finite interval is reduced to two possible extensions. This suggests relating these two types of extensions to the type of particle as being fermion or boson.

## 1. Introduction

Let $M$ be an $N$-dimensional $C^\infty$ Riemannian manifold, $x = (x^1,...,x^n)$ be a chart on $M$, except perhaps for a closed trivial subset, $g_{ij}$ be the metric tensor in terms of the given chart, and $(p_1,...,p_n)$ be the classical generalized momenta. The method of canonical quantization [1] is based on the assumption that associated with the classical canonical momenta $x^k, p_k (k = 1,...,n)$ obeying the usual Poisson bracket relations there exist quantum canonical observables $\hat{x}_k, \hat{p}_k$ satisfying the corresponding commutation relations. The operators $\hat{x}_k, \hat{p}_k$ act on the Hilbert space of complex valued square-integrable functions $L^2(M)$, and must be self-adjoint in their domain of definition. It is evident that the operator $\hat{x}_k$ defined in $L^2(M)$ by $\hat{x}^k \psi = x^k \psi$ on the domain

$$D(\hat{x}_k) = \{\psi \in AC_0(M) : \psi, x^k\psi \in L^2(M)\},$$

where $AC_0(M)$ denotes the set of absolutely continuous functions of compact support on M, is self-adjoint. The operator

$$\hat{p}_k = -i\hbar(\partial_k + \tfrac{1}{2} div \partial_k), \qquad (1)$$

where $\partial_k = \partial/\partial x^k$, defined on the domain

$$D_0(\hat{p}_k) = \{\psi \in AC_0(M) : \psi, \hat{p}_k\psi \in L^2(M)\} \qquad (2)$$

is symmetric [5]. The existence of a self-adjoint extension of the symmetric operator $\hat{p}_k$ depends on the range of $x^k$ as being a finite interval, the semi-axis, or the whole real line, in the following way [5].

(i) If the range of $x^k$ is a finite interval of $\Re$, $\hat{p}_k$ admits a one-parameter family of self-adjoint extensions.

(ii) If the range of $x^k$ is the semi-axis $(0,\infty)$, $\hat{p}_k$ is maximal symmetric and has no self-adjoint extensions.

(iii) If the range of $x^k$ is the whole real line, $\hat{p}_k$ is essentially self-adjoint, and has a unique self-adjoint extension, namely $\hat{p}_k^+$. (appendix A).

In this concern attention should be drawn to the important work of Mackey [3] who gave a coordinate free treatment of the problem of momentum quantization, proving that to every $C^\infty$ vector field $L$ in $M$ there corresponds a symmetric operator in $L^2(M)$:

$$\hat{P} = -i\hbar(L + \tfrac{1}{2} divL) \qquad (3)$$

with domain of definition

$$D_0(\hat{P}) = \{\psi \in L^2(M) : \psi \in C_0^\infty(M), \hat{P}\psi \in L^2(M)\}, \qquad (4)$$

and that $\hat{P}$ is essentially self-adjoint if and only if $L$ is complete. The latter result was also given [3] by Van Hove (1951). It is evident that the operator (1) is a special case of the operator (3) corresponding to $L = \partial_k$.

In this work we single out the class of incompressible momentum operators as exhibiting, to a great deal, Cartesian-like features, in the sense that incompressible momenta in a manifold $M$ resemble the Cartesian momenta $\hat{p}_x, \hat{p}_y, \hat{p}_z$ in a 3-Euclidean space with Cartesian coordinates $(x, y, z)$, which we denote by $E^3$. In fact the eigenfunctions of an incompressible momentum are almost a generalization of de Broglie waves, a feature which has an immediate consequence manifested in uniform boundedness of the wave functions and their square integrability on every bounded measurable subset of M. We also present an attempt to remove the absurdity associated with the existence of an infinite number of self-adjoint extensions of the operator (1) when the conjugate coordinate ranges in a finite interval.

## 2. Features of Incompressible Momenta

Let $L$ be a $C^1$ vector field in $M$. The vector field $L$ is said to be incompressible if $divL=0$. The notion of compressibility removing factors (CRFs for short) and incompressible momenta which were introduced in [7] are restated here.

*Definition* 1. A $C^1$ function $\mu : M \to \Re$ is a CRF of a $C^1$ vector field $L$ in $M$ if the vector field $\mu L$ is incompressible.

*Definition* 2. The operator $\hat{P} = -iL$ in $L^2(M)$ is described as incompressible if the vector field $L$ is incompressible.

We shall be concerned here with incompressible operators of the form

$$\hat{P}_k = -i\mu_k \partial_k \equiv \mu_k q_k \qquad (5)$$

where $\hbar$ is taken equal to 1, $\mu^k$ is a CRF of the field $\partial_k$, and $q_k \equiv -i\partial_k$.

We list here the analogy and deference between an incompressible momentum in $M$, say $\hat{P}_i$, and the Cartesian momentum $\hat{p}_x$ in $E^3$.

(i) The eigenfunctions of the operator $\hat{P}_1$, namely the solutions of the eigenequation $-i\mu^1 \partial \psi / \partial x^1 = P_1 \psi$, are easily found to be

$$\psi_{P_1} = f^{(1)} \exp(iP_1 \int dx^1 / \mu^1). \qquad (6)$$

The function $f^{(1)}$ in (6), which is independent of $x^1$ and arbitrary, can be determined through adjoining to $\hat{P}_1$ a set of observables to obtain a maximal set. The eigenfunctions of $\hat{p}_x$ namely the de Broglie waves $\phi_{p_x}(x,y,z) = f(x,y)\exp(ip_x x)$ associate with it a probability density distribution of the coordinate $x$ that is independent of $x$. This property is also enjoyed by the incompressible operator $\hat{P}_1$, where by (6), $|\psi_{P_1}|^2 = |f^{(1)}|^2$ is also independent of $x^1$.

(ii) The set of observables $(\hat{p}_x, y, z)$ (spin is discarded) is maximal. Similarly the coordinates $x^j (j = 2,...,n)$ commute with $\hat{P}_1$ so that $\{\hat{P}_1, x^2,...,x^n\}$ is maximal. However, whereas $\{\hat{p}_x, \hat{p}_y, \hat{p}_z\}$ is maximal, there may not exists in (M,x) a corresponding maximal set of incompressible momenta. In fact the existence of such a set imposes on the determinant g of the metric tensor $g_{ij}$ in M to be of a specific form. This will be discussed in the next section.

(iii) It is clear that the operators $\hat{x}_k, \hat{P}_k$ are in general not canonically conjugate. However, in an important special case it is possible to construct a set of observables $X^1(x^1),..., X^n(x^n)$, which form together with the incompressible momenta (3) a set of canonically conjugate variables.

(iv) The generators of the symmetry group of a manifold, or Killing fields, are incompressible [9] and also are constant of free motion. The Cartesian momentum $\hat{p}_x$ for example is a constant of free motion, a fact that stems from being generated by the group of translation [6] of $E^3$ along ox. An incompressible momentum is not a constant of the free motion unless it is generated by a symmetry group of the manifold.

(v) The Hamiltonian operator of the free motion in M can be expressed in terms of n incompressible momentum operators of the form (5) as follows [Appendix A]:

$$\hat{H} = \frac{1}{2m} \sum \hat{P}_i \frac{g^{ij}}{\mu^i \mu^j} \hat{P}_j \qquad (7)$$

The above relation generalizes the familiar relation $\hat{H} = (2m)^{-1}(\hat{p}_x^2 + \hat{p}_y^2 + \hat{p}_z^2)$ in $E^3$.

## 3. A Complete Set of Incompressible Generalized Momenta

The CRF $\mu^k$ of the vector field $\partial_k$ satisfies the equation [7]

$$div(\mu^k \partial_k) = g^{-1/2} \frac{\partial}{\partial x^k}(\mu^k g^{1/2}) = 0 \qquad (8)$$

where $g = \det(g_{ij})$. We note that no summation over repeated indices is implied through out this work and that we shall use comma followed by an index to denote differentiation with respect to a coordinate (for example $\psi_{,k} = \partial \psi / dx^k$). Equation (8) yields $\mu^k \sqrt{g} = I^k$, where $I^k$ is an arbitrary function that is independent of $x^k$. If each $\mu^k$ $(k = 1,...,n)$, is to depend on $x^k$ alone, then $\sqrt{g}$ has to assume the form

$$\sqrt{g} = g^1(x^1)....g^n(x^n) = 1/\mu^1(x^1)...\mu^n(x^n), \tag{9}$$

where we have set $\mu^k(x^k) = [g^k(x^k)]^{-1}$. Therefore if $g$ can be factorized to products as prescribed by the last relation then for each momentum operator $\hat{P}_k$ $(k = 1,....,n)$ we can choose a CRF $\mu^k(x^k)$ which depends only on the coordinate $x^k$ (it may be a constant in a special case). A momentum of the form $\hat{P}_k = -i\mu^k(x^k)\partial_k$ will be called an incompressible generalized momentum (IGM for short). Assuming that the metric's determinant $g$ can be factorized as given by (9) the following set of IGM observables

$$\hat{P}_k = -i\mu^k \partial_k \quad k \in [1,n] \tag{10}$$

is maximal (spin is discarded). A common eigenfunction of this set belonging to the eigenvalue $(P_1,....,P_n)$ has the form

$$|P_1....P_n> = ce^{iP_1 \int \frac{dx^1}{\mu^1} + .....+ iP_n \int \frac{dx^n}{\mu^n}} \tag{11}$$

where $c$ is a normalization constant. Every eigenfunction of the form (11) determines a uniform probability density distribution of the coordinates, for $<P_1....P_n | P_1....P_n> = |c|^2$; and hence the probability of finding the particle in a volume element $dV$ about a point is independent of its position.

We define now $n$ functions of the coordinate, which will form together with the set of IGM observables a set of canonically conjugate observables. Let

$$X^k(x^k) = \int_{a^k}^{x^k} \frac{dt}{\mu^k(t)} + X^k(a^k) \quad k = 1,...,n \tag{12}$$

where we assume that $a^k$ is any point in the range of the coordinate $x^k$. It is evident that

$$[\hat{X}^k, \hat{P}_j] = i\delta_{kj} \tag{13}$$

and hence [10]

$$\Delta X^k \Delta P_k \geq \tfrac{1}{2} \tag{14}$$

By (12) the eigenfunctions (11) are written as

$$|P_1.....P_n> = ce^{i(X^1 P_1 +....+ X^n P_n)} \tag{15}$$

which is formally the same as de Broglie wave.

Now if $g$ can be factorized as in (9), the expectation value of an IGM operator, say $\hat{P}_1$, when the system is in an arbitrary state $\psi \in D(\hat{P}_1)$ is given by

$$<\hat{P}_1>_\psi = \int g_2 dx^2 ... g_n dx^n \int \psi^*(-i\partial\psi/\partial x^1)dx^1.$$

This shows that the expectation value of an incompressible momentum $\hat{P}_k$ is equal to the expectation value of the operator $\hat{q}_k = -i\partial_k$ calculated as if the conjugate coordinate $x^k$ were Cartesian. In other words, the expectation value of $\hat{P}_1 = -i\mu^1 \partial/\partial x^1$ in a state $\psi$ is the same as that of the operator $\hat{q}_1 = -i\partial/\partial x^1$ considered as if acting in another Hilbert space, namely that formed by complex valued functions defined on M which are square integrable with respect to the volume element $dx^1.g_2 dx^2 ...... g_n dx^n$.

## 4. Self-Adjoint Extensions

The following theorem specifies the cases in which the symmetric operator $\hat{Q}_k = -i\mu^k \partial_k$ defined in $L^2(M)$ on the set $AC_0(M)$ of absolutely continuous functions with compact support admits a self-adjoint extension.

**Theorem 1**: (A) The incompressible momentum $\hat{Q}_k$ defined in $L^2(M)$ on the domain

$$D(Q_k) = \{\psi \in L^2(M) : \psi \in AC_0(M), \mu^k \psi_{,k} \in L^2(M)\}$$

is symmetric.

(B) The adjoint operator $Q_k^+$ is defined by $Q_k^+ \psi = -i\mu^k(x^k)\partial_k \psi$ on the domain

$$D(Q_k^+) = \{\psi \in L^2(M) : \psi \in AC(M), \mu^k \psi_{,k} \in L^2(M)\}.$$

(C) The existence of a self-adjoint extension $\hat{P}_k$ of $Q_k$ depends on the range $(a,b)$ of the coordinate $x^k$:

(i) If $(a,b) = \Re,$ the operator $\hat{P}_k \equiv Q_k^+$ is self-adjoint.

(ii) If $(a,b) = (a,\infty),$ the restriction $\hat{P}_k$ of the operator $Q_k^+$ to the domain

$$D(\hat{P}_k) = \{\psi \in D(Q_k^+) : \lim_{x^k \to 0} \psi(x^k) = 0\}$$

is maximal symmetric; it has no self-adjoint extensions.

(iii) If $(a,b)$ is finite, $Q_k$ admits a one-parameter family $\hat{P}_k^\alpha, \alpha \in [0, 2\pi)$ of self-adjoint extensions. A typical extension is the restriction of $Q_k^+$ to the domain

$$D(\hat{P}_k^\alpha) = \{\psi \in D(Q_k^+) : \lim_{x^k \to a} \psi(x) = e^{i\alpha} \lim_{x^k \to b} \psi(x)\}.$$

The mathematical result (ciii) implies that if $a < x^k < b$ where $a$ and $b$ are finite and if $\lim_{x^k \to a} \psi(x), \lim_{x^k \to b} \psi(x)$ correspond to distinct points in $M$, then there exist many self-adjoint extensions $\hat{P}_k^\alpha$ of $Q_k$, one for each value of $\alpha \in [0, 2\pi)$. Utilizing $(6)$ and the boundary condition in (ciii) we deduce that each $\hat{P}_k^\alpha$ has a discrete spectrum

$$\{P_{kj}^\alpha = (\int_a^b \frac{dx^k}{\mu^k})^{-1}(2\pi j - \alpha) : j \in Z\}$$

with the corresponding eigenfunctions

$$\{\psi_{kj}^\alpha = f^{(k)} \exp(iP_{kj}^\alpha \int_a^{x^k} \frac{dx^k}{\mu^k}), j \in Z\}.$$

By (12) we see that the spectrum of $\hat{P}_k^\alpha$ and the corresponding eigenfunctions are

$$P_{kj}^\alpha = [X^k(b^k)]^{-1}(2j\pi - \alpha), \quad \psi_{kj}^\alpha = f^{(k)} \exp(iP_{kj}^\alpha X^k(x^k)) \quad (j = 1,2,3,....)$$

However we shall interpret the result (ciii) physically as implying that the phase of the wave function at one end point of the interval is advancing by $\alpha$ from its value at the other end. This interpretation is applicable to each end, and hence $\psi(a) = e^{i2\alpha}\psi(a)$ which yields $\alpha = 0\ or\ \pi$. This is equivalent to postulating that $\psi(b) = \pm\psi(a)$. We have initially, therefore, two possible extensions $\hat{P}_k^0$ and $\hat{P}_k^\pi$ with the corresponding spectra

$$P_{kj}^0 = (\int_a^b \frac{dx^k}{\mu^k})^{-1} 2j\pi, \quad P_{kj}^\pi = (\int_a^b \frac{dx^k}{\mu^k})^{-1}(2j-1)\pi \quad (j \in Z) \qquad (16)$$

It is clear that when $\lim_{x^k \to a} \psi(x), \lim_{x^k \to b} \psi(x)$ correspond to the same point then only the extension $\hat{P}_k^0$ is admissible.

We may interpret the existence of two types of extension as indicating that there exist two types of systems characterized by the type of boundary condition on their wave functions. It is natural then to suppose that these two types are bosons and fermions systems to which the self-adjoint extensions $\hat{P}_k^0$ and $\hat{P}_k^\pi$ correspond respectively.

We finally draw attention to the important work by Wan and others [4] in which a new approach to momentum quantization is developed. In this approach the set of quantizable classical momenta is enlarged to comprise those which are representable by maximal symmetric operators. However the non-uniqueness problem intimate to a finite range of a coordinate persists, and every $\hat{P}^\alpha$ accordingly is a legitimate momentum operator.

## 5. Example

Let $M = \{(x, y, z) \in \Re^3 : (x^2 + y^2 + z^2)^{1/2} < a\}$ be a subset of the 3-dimensional Euclidean space and $(r, \phi, \theta)$ be the system of spherical coordinates in M in which the metric is $ds^2 = dr^2 + r^2(d\theta^2 + \sin^2\theta\, d\phi^2)$. It is evident that the momentum operators

$$\hat{P}_r = -ir^{-2}\partial_r,\ \hat{P}_\theta = -i(\sin\theta)^{-1}\partial_\theta,\ \hat{L}_z = -i\partial_\phi$$

are incompressible and mutually commutative. The functions

$$\mu_r = r^{-2},\ \mu_\theta = (\sin\theta)^{-1},\ \mu_\phi = 1$$

are the CRFs of the fields $\partial_r, \partial_\theta, \partial_\phi$ respectively.

It is well known that the operator $\hat{L}_z$ defined on the domain

$$D(\hat{L}_z) = \{\psi \in AC(M) : \psi, \psi_{,\phi} \in L^2(M), \psi(r,\theta,0) = \psi(r,\theta,0)\}$$

is self-adjoint [2]. The proof of this statement follows also from theorem 1 (ciii), which also shows that $Sp(L_z) = \{m\hbar : m \in Z\}$ with the corresponding eigenfunctions $\psi_m(\phi) = (2\pi)^{-1/2} e^{im\phi}, (m \in Z)$, where $\psi_m(\phi)$ denotes only the $\phi$-part of the wave function.

By theorem 1, and since the coordinate $\theta$ ranges over the finite interval $(0, \pi)$, each operator $\hat{P}_\theta^\alpha$, where $\alpha = 0\ or\ \pi$, whose domain of definition is

$$D(\hat{P}_\theta^\alpha) = \{\psi \in L^2(M) : \psi \in AC(M), \hat{P}_\theta^\alpha\psi \in L^2(M), \psi(r,0,\phi) = \psi(r,\pi,\phi)\}$$

is self-adjoint. By (15) the spectrum and the corresponding eigenfunctions of each of these operators are given by

$$\psi_k(\theta) = 2^{-1/2} \exp\{-i(k\pi - \alpha/2)\cos\theta] \qquad (k \in Z)$$
$$Sp(\hat{P}_\theta^\alpha) = \{(k\pi - \alpha/2) : k \in Z\}.$$

We shall see soon that the value $\alpha = \pi$ is discarded. Since $0<r<a$, there exists two possible radial momenta $\hat{P}_r^\beta$ ($\beta = 0\, or\, 1$), with domains

$$D(\hat{P}_r^\beta) = \{\psi \in L^2(M) : \psi \in AC(M), \hat{P}_r^\beta \psi \in L^2(M) : \psi(0,\theta,\phi) = e^{i\beta}\psi(a,\theta,\phi)\}$$

It follows that for either value $\beta = \pm 1$ we must have $\psi(a,0,\phi) = \psi(a,\pi,\phi)$. This result shows also that $\hat{P}_\theta^\pi$ is not admissible, whereas only $\hat{P}_\theta^0$ is admissible when taken together with $\hat{P}_r^\pi$ as members of a complete set of observables in M. The IGM observables $\{\hat{P}_r^\beta, \hat{P}_\theta^0, \hat{L}_z\}$ we have discussed have non-degenerate discrete spectra with the common set of orthogonal eigenfunctions

$$|mjk> = \sqrt{\frac{3}{4\pi a^3}} \exp i[m\phi + k\pi(1-\cos\theta) + a^{-3}(2j\pi - \beta)r^3] \quad (m,k,j \in Z),$$

where $\beta$ is either $0$ or $\pi$.

By equation (7) the Hamiltonian operator can be expressed in terms of the previous set of IGM observables as follows:

$$\hat{H} = \frac{1}{2m}[\hat{P}_r r^4 \hat{P}_r + \frac{1}{r^2}\hat{P}_\theta \sin^2\theta\, \hat{P}_\theta + \frac{1}{r^2 \sin^2\theta}\hat{L}_z^2].$$

## Appendix A: Proof of Theorem 1

Let $(a,b)$ denote a finite interval, or a semi-axis $(a,\infty)$, or the whole real line $\Re$. In order to simplify notations we prove the theorem for $Q_1 = -i\mu^1(x^1)\partial_1$.

(A) For every $\psi, \phi \in D(Q_1)$ we have

$$(\psi, Q_1\phi) = \int_a^b dx^1 \int \psi^*(-i\mu^1\phi_{,1})\sqrt{g}\,dx^2.....dx^n. \qquad (A1)$$

The second integral extends over the ranges of the coordinates $(x^2,....,x^n)$. Integrating by parts with respect to the coordinate $x^1$, and recalling that $(\mu^1\sqrt{g})_{,1} = 0$, we obtain

$$(\psi, Q_1\phi) = -i\int [\psi^* \phi \sqrt{g}]_a^b dx^2.....dx^n + \int_a^b dx^1 \int (-i\mu^1 \psi_{,1})^* \phi \sqrt{g} dx^2.....dx^n \quad (A2)$$

The first integral on the right hand side of the equality vanishes on the account of the boundary conditions

$$\lim_{x^1 \to a} \phi(x^1) = \lim_{x^1 \to b} \phi(x^1) = 0 \quad (A3)$$

Hence $(\psi, Q_1\phi) = (Q_1\psi, \phi) \; \forall \psi, \phi \in D(Q_1)$, and $Q_1$ is symmetric.

(B) The domain of $Q_1^+$ is composed of all functions $\psi \in L^2(M)$ for which there exists $\eta \in L^2(M)$ such that

$$(\psi, Q_1\phi) = (\eta, \phi) \qquad \forall \psi, \phi \in D(Q_1). \quad (A4)$$

The coordinates $x^k$ $(k \neq 1)$ can be ignored without affecting the validity of our argument, or equivalently we may treat the problem as if it were in one dimension

$$(\psi, Q_1\phi) = \int_a^b \psi^* (-i\mu^1 \phi_{,1}) \sqrt{g} \, dx^1 \quad (A5)$$

$$(\eta, \phi) = \int_a^b \eta^* \phi \sqrt{g} dx^1 = -\int_a^b \phi_{,1} [\int_0^{x^1} \eta(t, x^2,..., x^n) \sqrt{g(t, x^2,..., x^n)} dt + c]^* dx^1 \quad (A6)$$

The last equality can be verified by integrating the right hand-side by parts and using (A3). Equating the right hand-sides of (A5) and (A6) yields

$$\int_a^b \phi_{,1} [i\mu^1 \sqrt{g} \psi + \int \eta(t) \sqrt{g} dt + c_1]^* dx^1 = 0 \quad \forall \phi \in D(Q_1)$$

Since $D(Q_1)$ is dense [2] in $L^2(M)$ we deduce that

$$-i\mu^1 \sqrt{g} \psi = \int \eta(t) \sqrt{g} dt + c_2$$

Differentiation with respect to $x^1$, and noting that $(\mu^1 \sqrt{g})_{,1} = 0$, yields

$-i\mu \sqrt{g} \partial_1 \psi = \eta \sqrt{g}$. It follows that

$$\eta = -i\mu^1 \partial_1 \psi, \quad Q_1^+ = -i\mu^1 \partial_1$$

and that the functions $\psi$ which form $D(Q_1^+)$ need not to fulfill (A3); it needs only to be absolutely continuous.

(C) The proof is similar to that given by Akhiezer and Glazmann [2].

## Appendix B: The Hamiltonian of the free motion in terms of IGM observables

We first note that the operator $\hat{q}_k = -i\partial_k$ defined in $L^2(M)$ on the domain

$$D(\hat{q}_k) = \{\psi \in L^2(M) : \psi \in AC_0(M), \partial_k\psi \in L^2(M)\}$$

is not symmetric; its adjoint is $\hat{q}_k^+ = -i(\partial_k + div\partial_k) = -ig^{-1/2}\partial_k g^{1/2}$ with domain $\{\psi \in AC(M) : \psi, \hat{q}_k^+\psi \in L^2(M)\}$. The proof of the last statements is systematic (see [2] for similar proofs). The operator $\hat{P}_k = \frac{1}{2}(\hat{q}_k + \hat{q}_k^+) = -i(\partial_k + \frac{1}{2}div\partial_k)$ defined on the domain (2) is symmetric. The existence of a self-adjoint extension of $\hat{P}_k$ depends on the range of the coordinate [5]. Leaving aside the problem of domains we have

$$\hat{H} = -(2m)^{-1}\sum_{i,j} g^{-1/2}\partial_i g^{1/2} g^{ij}\partial_j = (2m)^{-1}\sum_{i,j}\hat{q}_i^+ g^{ij}\hat{q}_j.$$

By (5) we have $\hat{q}_k = (1/\mu^k)\hat{P}_k, \hat{q}_k^+ = \hat{P}_k(1/\mu_k),$ and hence

$$\hat{H} = (2m)^{-1}\sum_{ij}\hat{P}_i \frac{g^{ij}}{\mu^i\mu^j}\hat{P}_j.$$

In the case when the coordinates are orthogonal the last relation takes the form

$$\hat{H} = (2m)^{-1}\sum_i \hat{P}_i \frac{g^{ii}}{(\mu^i)^2}\hat{P}_j.$$